\documentclass[aps,prc,twocolumn,amsmath,amssymb,superscriptaddress]{revtex4-1}
\usepackage{graphicx}
\usepackage{hyperref}
\usepackage{color}
\newcommand{\be}{\begin{equation}}
\newcommand{\ee}{\end{equation}}
\newcommand{\beq}{\begin{eqnarray}}
\newcommand{\eeq}{\end{eqnarray}}
\newcommand{\ba}{\begin{array}}
\newcommand{\ea}{\end{array}}

\begin{document}

\title{$\rho$-Meson Nucleon Scattering Length from CLAS Threshold Photoproduction Measurements}

\date{\today}

\author{\mbox{Igor~I.~Strakovsky}}
\altaffiliation{Corresponding author: \texttt{igor@gwu.edu}}
\affiliation{Institute for Nuclear Studies, Department of Physics, The George Washington University, Washington, DC 20052, USA}

\author{\mbox{Evgeny~L.~Isupov}}
\altaffiliation{Corresponding author: \texttt{isupov@jlab.org}}
\affiliation{Skobeltsyn Institute of Nuclear Physics, Lomonosov Moscow State University, 119234 Moscow, Russia}

\author{\mbox{Victor~Mokeev}}
\altaffiliation{Corresponding author: \texttt{mokeev@jlab.org}}
\affiliation{Thomas Jefferson National Accelerator Facility, Newport News, VA 23606, USA}

\author{\mbox{Axel~Schmidt}}
\altaffiliation{Corresponding author: \texttt{axelschmidt@gwu.edu}}
\affiliation{Institute for Nuclear Studies, Department of Physics, The George Washington University, Washington, DC 20052, USA}

\noaffiliation

\begin{abstract}
Extending our study of the vector meson-nucleon scattering lengths (summary is given in Ref.~\cite{Strakovsky:2021vyk}), we are focusing on the $\rho$-meson case using recent CLAS threshold data for the reaction $\gamma p \to \rho p$ within the meson–baryon reaction model~\cite{CLAS:2018drk}. The total $\sigma_t(\gamma p\to \rho p)$ and $\sigma_t(\gamma p\to \omega p)$ cross sections are close  
below the momentum of the vector meson in CM $q = 0.2~\mathrm{GeV/c}$. Then the $\rho p$ photoproduction cross section grows rapidly.
Our result for $\rho N$ scattering length is a factor of 4 smaller than the size of the hadron and the phenomenological determination of the $\omega$ nucleon scattering length using threshold photoproduction cross sections. The observed difference between $\rho N$ and $\omega N$ scattering lengths is of interest for further understanding within the hadron structure theory.
\end{abstract}

\maketitle

\section{Introduction}
\label{sec:intro} 
\begin{figure}[htb!]
\centering
{
    \includegraphics[width=0.9\columnwidth]{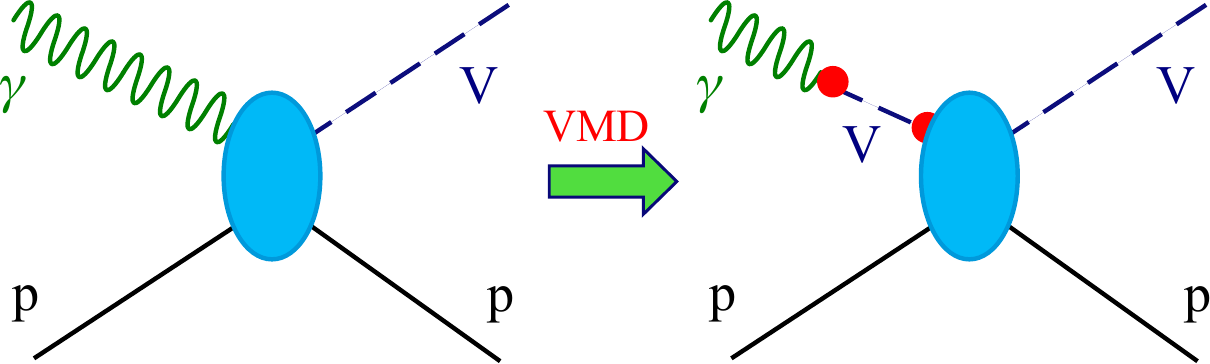}
}

\centerline{\parbox{0.4\textwidth}{
\caption[] {\protect\small
 Schematic diagrams of vector-meson photoproduction (left) and the VMD model (right) in the energy region at threshold experiments.
} 
\label{fig:fig1} } }
\end{figure}

There are no vector meson (V) beams, so experiments using modern electromagnetic (EM) facilities attempt to access vector meson nucleon (VN) interactions via EM production reactions $ep\to e^\prime Vp$. These reactions can be measured with very high precision, because V's have the same quantum numbers space (P) and charge (C) parities as the photon: $J^{PC} = 1^{-~-}$. 
It allows us to apply Sakurai’s Vector Meson Dominance (VMD) model, assuming that a real photon can fluctuate into a virtual V, which subsequently scatters off a target nucleon~\cite{Sakurai:1960ju} (Fig.~\ref{fig:fig1}). Comparisons of different Vs allow the systematic study of the general properties of the VN interaction.

Our previous study focused on 4 Vs ($\omega$, $\phi$, $J/\psi(1S)$, and $\Upsilon(1S)$) from the $q\bar{q}$ nonet, the widths of which are narrow enough to study meson photoproduction at threshold, and where data and quasi-data, \textit{i.e.}, theoretical predictions, were available~\cite{Strakovsky:2021vyk}.
These studies were focused on the systematics of the VN scattering length (SL). To avoid the problem of a broad resonance width at threshold, we did not consider the $\rho$-meson. Furthermore, we ignored, for example, $\psi^\prime(2S)$, whose difference from the $1S$ state comes purely from the radial wave function. Unfortunately, we cannot consider Quarkonium beyond the $\Upsilon$ ($b\bar{b}$). The problem is that the Toponium ($T(1S)$), whose quark content is $t\bar{t}$, does not exist, given that the $t$-quark decays faster than the meson formation time.

Occasionally, the $\rho$-meson is considered as the gauge boson of a broken hidden local chiral symmetry, which, however, is distinct from the broken global chiral symmetry of which the pion is the respective Goldstone boson~\cite{Georgi:1989xy}. Moreover, the non-strange vector mesons $\rho$, $\omega$, and $\phi$ play a central role in Sakurai’s VMD model, where they account for the (virtually admixed) hadronic part of the photon by which the photon couples to hadronic matter. The applicability of the vector meson dominance (VMD) model was recently explored in Ref.~\cite{Xu:2021mju}. An important conclusion from these studies is that the VMD approximation is justified primarily for the photoproduction of light vector mesons, namely the $\rho$ and $\omega$ mesons.

The $\rho$- and $\omega$-mesons were detected first in 1961 at LBL~\cite{Maglich:1961rtx}. Remarkably, the $\omega(782)$ meson, being the isoscalar partner of the $\rho$-meson, is a rather long-lived object with a small total width of about $\Gamma_\omega \sim 9~\mathrm{MeV}$~\cite{ParticleDataGroup:2024cfk}. The increase in lifetime/decrease in decay width by about a factor of 20 is 
an effect of spin-parity conservation laws, inhibiting decays of an isoscalar vector meson into two pions. An even more extreme case is the $\phi(1020)$ meson, which is the singlet partner of the $\omega$-meson.

This paper focuses on the study of the $\rho$-meson proton scattering length. The evaluation of $\gamma p \to \rho p$ cross sections is challenging because of the large hadronic decay width of the $\rho$ meson, $\Gamma_\rho \approx 147~\mathrm{MeV}$. As a result, the $\rho$-meson must be treated as a resonance decaying into the final $\pi^+\pi^-$ state. To extract the $\rho$-meson scattering length, it is necessary to explore the near-threshold region of $\rho p$ photoproduction. At $W < 1.8~\mathrm{GeV}$, the $\rho p$ contribution is smaller than other mechanisms contributing to $\pi^+\pi^-p$ photoproduction. Therefore, the extraction of $\gamma p \rightarrow \rho p$ cross sections requires amplitude analyses of the experimentally accessible $\gamma p \rightarrow \pi^+\pi^-p^\prime$ photoproduction cross sections.

The CLAS Collaboration measurements of the $\gamma p \to \pi^+\pi^-p^\prime$ reaction~\cite{CLAS:2018drk} provided, for the first time, nine independent one-fold differential cross sections in each $W$ bin across the range $1.6 < W < 2.0~\mathrm{GeV}$. Amplitude analyses of these results allowed us to determine the integrated $\gamma p \to \rho p$ cross sections, which were then used as input for the evaluation of the $\rho$-meson scattering length.

\section{Extraction of Integrated $\gamma p \rightarrow \rho p$ Photoproduction Cross Sections}
\label{rhop_sect}
The $\gamma p \rightarrow \rho p$ photoproduction cross sections were obtained from an analysis of nine one-fold differential $\pi^+\pi^-p$ photoproduction cross sections measured with the CLAS detector. The analysis was carried out in bins of the invariant mass of the three-body final hadronic system, using $25~\mathrm{MeV}$ bin widths and covering the $W$ range from 1.6 to $2.0~\mathrm{GeV}$~\cite{CLAS:2018drk}. Each of the nine one-fold differential $\pi^+\pi^-p$ cross sections was evaluated as an integral of a common five-fold differential cross section, which fully determined the kinematics of the final $\pi^+\pi^-p$ state, over different sets of four variables.

Defining $M_{\pi^+p}$, $M_{\pi^-p}$, and $M_{\pi^+\pi^-}$ as the invariant masses of the three possible two-hadron subsystems in the final state, we adopt the following choice for evaluating the five-fold differential cross section:
\begin{equation}
    d^5\tau = dM_{\pi^+p} \, dM_{\pi^+\pi^-} \, d\Omega_{\pi^-} \, d\alpha_{[\pi^-p][\pi^+p^\prime]} \>,
\label{eq:eq0}
\end{equation}
where $\Omega_{\pi^-}$ is the solid angle of the $\pi^-$ in the final-state center-of-mass (CM) frame, specified by its polar angle $\theta_{\pi^-}$ and azimuthal angle $\phi_{\pi^-}$. The variable $\alpha_{[\pi^-p][\pi^+p^\prime]}$ denotes the angle between two planes: the first defined by the momenta of the $\pi^-$ and the initial proton $p$, and the second by the momenta of the $\pi^+$ and the final proton $p^\prime$, measured about the axis given by the $\pi^-$ momentum.

The one-fold differential cross sections consist of:
\begin{itemize}
\item invariant-mass distributions of the three two-particle subsystems: $d\sigma/dM_{\pi^+\pi^-}$, $d\sigma/dM_{\pi^+p}$, and $d\sigma/dM_{\pi^-p}$;
\item distributions over CM polar angles of the three final-state particles: $d\sigma/(\sin\theta_{\pi^-}d\theta_{\pi^-})$, $d\sigma/(\sin\theta_{\pi^+}d\theta_{\pi^+})$, and $d\sigma/(\sin\theta_{p^\prime}d\theta_{p^\prime})$;
\item distributions over the three $\alpha$-angles defined in the CM frame: $d\sigma/d\alpha_{[\pi^-p][\pi^+p^\prime]}$, $d\sigma/d\alpha_{[\pi^+p][\pi^-p^\prime]}$, and $d\sigma/d\alpha_{[\pi^+\pi^-][pp^\prime]}$, where the last two are defined analogously to $d\sigma/d\alpha_{[\pi^-p][\pi^+p^\prime]}$ described above.
\end{itemize}
Further details on the extraction of these differential cross sections from the data can be found in~\cite{CLAS:2018drk, Mokeev:2012vsa}.

In the evaluation of $\rho$ meson scattering length, we are using the integrated $\gamma p \rightarrow \rho p$ photoproduction cross section deduced from analyses of the aforementioned nine one-fold differential cross sections~\cite{CLAS:2018drk}. These differential cross sections were analyzed within the meson–baryon reaction model (JM), developed in collaboration between Jefferson Lab and Moscow State University. 

\begin{figure*}[htp]
\begin{center}
\includegraphics[width=12.8cm]{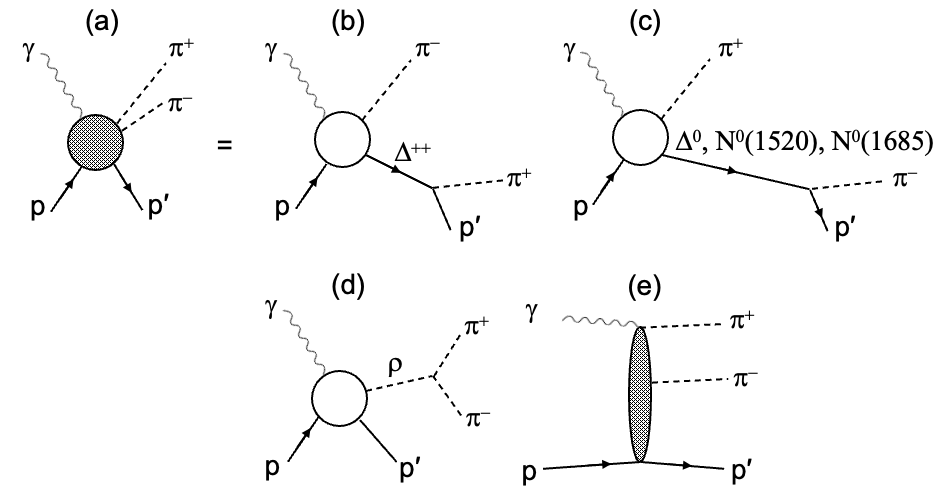}
\vspace{-3mm}
\caption{The $\gamma p\to  \pi^+\pi^- p'$ photoproduction mechanisms incorporated into the JM model~\cite{CLAS:2018drk, Mokeev:2012vsa}: 
a) full amplitude; 
b) $\pi^- \Delta^{++}$; 
c) $\pi^+ \Delta^0$, $\pi^+ N^0(1520)3/2^-$, and $\pi^+N^0(1680)5/2^+$ sub-channels; 
d) $\rho p$ sub-channel; and
e) direct $2 \pi$ mechanisms.}
\label{jmmech}
\end{center}
\end{figure*}
The JM model incorporates contributions from the $\pi^-\Delta^{++}$, $\rho p$, $\pi^+\Delta^0$, $\pi^+ N(1520)3/2^+$, and $\pi^+ N(1680)5/2^+$ meson–baryon channels. In addition, direct $2\pi$ photoproduction mechanisms are included, which produce the final $\pi^+\pi^-p$ state without intermediate unstable hadrons. The three-body amplitudes shown in Fig.~\ref{jmmech} account for the production, propagation, and decay of unstable mesons and baryons into final states composed of stable hadrons. The propagation of unstable hadrons is described by Breit–Wigner propagators, which include their decay widths and their dependence on the invariant masses of the final-state hadron pairs. Within the JM approach, the finite decay width of the $\rho$ meson is explicitly taken into account together with the contributions from the other mechanisms mentioned above relevant for $\pi^+\pi^-p$ photoproduction at $W < 2.0~\mathrm{GeV}$.

Nucleon resonances contribute through the $\pi^-\Delta^{++}$, $\rho p$, and $\pi^+\Delta^0$ channels. For the evaluation of the $\gamma p \rightarrow \rho p$ cross sections, we use the resonance parameters determined from fits to the CLAS $\pi^+\pi^-p$ photoproduction data~\cite{CLAS:2018drk}.

To describe the non-resonant amplitudes in the $\rho p$ channel within the region $W < 2.0~\mathrm{GeV}$, we adopted the simple ansatz developed in Ref.~\cite{Soding:1966}, which includes only $t$-channel exchanges parameterized by an exponential propagator. The magnitudes of the non-resonant $\rho p$ photoproduction amplitudes were fit to the data on the nine one-fold differential cross sections in each $W$ bin. We found that for squared momentum transfer between the initial and final proton, $t_{pp^\prime} < 1.0~\mathrm{GeV^2}$, this ansatz provides a satisfactory description of the data. In the region $t_{pp^\prime} > 1.0~\mathrm{GeV^2}$, the resonant contributions to the $\rho p$ channel dominate over the non-resonant background. Therefore, this parameterization of the non-resonant amplitudes is appropriate for use in the region $W < 2.0~\mathrm{GeV}$.

Further details on the evaluation of the meson–baryon diagrams shown in Fig.~\ref{jmmech} can be found in Refs.~\cite {CLAS:2018drk, Mokeev:2012vsa, Mokeev:2008iw, Ripani:2000va, Burkert:2007kn}.

A good description of the nine one-fold differential $\pi^+\pi^-p$ photoproduction cross sections in the range $1.6 < W < 2.0~\mathrm{GeV}$ has been achieved within the JM model, with $\chi^2/\text{d.p.} < 1.5$. Representative examples of the description of these cross sections can be found in Ref.~\cite {CLAS:2018drk}. The integrated $\gamma p \rightarrow \rho p$ cross sections were computed as a function of $W$, accounting only for the contributions from the diagrams shown in Fig.~\ref{jmmech}(d). They were further evaluated as a function of the absolute value of the $\rho$-meson three-momentum, as shown in Fig.~\ref{fig:fig2}.

The uncertainties were determined from the fit to the nine one-fold differential cross sections. They include both the experimental data uncertainties and those associated with the JM model resonance and non-resonance parameters. The uncertainties of the experimental data on nine one-fold differential cross sections~\cite{CLAS:2018drk} are given as the quadratic sum of the statistical and kinematically dependent contributions, with the latter being dominant 
(Table~\ref{tbl:tab0}).
\begin{table}[htb!]
\centering \protect\caption{{The first column showed invariant mass of the final hadron system, the second column showed the momentum of the $\rho$-meson in the CM frame, and the third column showed total cross section of the reaction $\gamma p \to \rho p$ with uncertainties extracted from CLAS measurements using JM model~\cite{CLAS:2018drk}.}
}
\vspace{2mm}
{%
\begin{tabular}{|c|c|c|}
\hline
  W       & q$_{min}$ & $\sigma_t$ \tabularnewline
(GeV)     &  (MeV/c)  & ($\mu b$)  \tabularnewline
\hline
1.737     & 143       &  5.910 $\pm$ 0.692 \tabularnewline
1.762     & 205       &  9.600 $\pm$ 1.216 \tabularnewline
1.787     & 253       & 11.840 $\pm$ 1.372 \tabularnewline
1.812     & 294       & 13.270 $\pm$ 1.472 \tabularnewline
1.837     & 330       & 14.090 $\pm$ 1.459 \tabularnewline
1.862     & 363       & 16.260 $\pm$ 1.636 \tabularnewline
1.887     & 394       & 18.710 $\pm$ 1.929 \tabularnewline
1.912     & 423       & 18.890 $\pm$ 1.990 \tabularnewline
1.937     & 450       & 18.240 $\pm$ 1.903 \tabularnewline
1.962     & 477       & 18.440 $\pm$ 2.009 \tabularnewline
1.987     & 502       & 16.720 $\pm$ 1.789 \tabularnewline
\hline
\end{tabular}} \label{tbl:tab0}
\end{table}

\section{$\rho$-Meson Scattering Length: Phenomenological analysis}
\label{rho-meson}
For evaluation of the absolute value of VN SL, we apply the VMD approach that links near-threshold photoproduction total cross sections of $\gamma p \to Vp$ and elastic $Vp \to Vp$~\cite{Strakovsky:2014wja, Strakovsky:2021vyk}.
To avoid theoretical uncertainties, we do not 
(i) determine the sign of SL, 
(ii) separate the Re and Im parts of SL, nor 
(iii) extract total angular momentum 1/2 and 3/2 contributions for $VN$ system. The limitations of VMD are listed in Refs.~\cite{Xu:2021mju, Kopeliovich:2017jpy, Boreskov:1976dj}.

Traditionally, the total cross section, $\sigma_t$, behavior of near-threshold binary inelastic reaction $ab\to cd$ with $m_a + M_b < m_c + M_d$ is described as a series of odd powers in the momentum of the meson in the CM frame, $q$: 
\begin{equation}
    \sigma_t = a~q + b~q^3 + c~q^5
    \>,
\label{eq:eq1}
\end{equation}
where the linear term is determined by two independent $S$-waves with total spin 1/2 and/or 3/2. Contributions to the cubic term come from both $P$-wave amplitudes and the $W$ dependence of $S$-wave amplitudes, and the fifth-order term arises from $D$-waves and the $W$ dependencies of $S$- and $P$-waves.

Finally, one can express the absolute value of VN SL, $|\alpha_{Vp}|$, as a product of 
the hadronic factor:
\begin{equation}
    h_{Vp} = \sqrt{a}
    \>
\label{eq:eq3}
\end{equation}
and pure EM VMD-motivated kinematic factor:
\begin{equation}
    B_V^2 = \frac{\alpha~m_V~k}{12~\pi ~\Gamma(V\to e^+e^-)}
    \>,
\label{eq:eq2}
\end{equation}
where the linear term $a$ comes from the best fit to the $\sigma_t$ of the V photoproduction (Eq.~(\ref{eq:eq1})),
$\alpha$ is the fine-structure constant,
$m_V$ is the mass of V,
$k$ is the photon CM momentum, and
$\Gamma(V\to e^+e^-)$ is the V partial width to 
$e^+e^-$~\cite{ParticleDataGroup:2024cfk}.

That absolute value SL is determined by the interplay of strong (hadronic) and EM dynamics as:
\begin{equation}
    |\alpha_{Vp}| = h_{Vp}~B_V
    \>.
\label{eq:eq4}
\end{equation}

\begin{figure}[htb!]
\centering
{
    \includegraphics[width=0.47\textwidth,keepaspectratio]{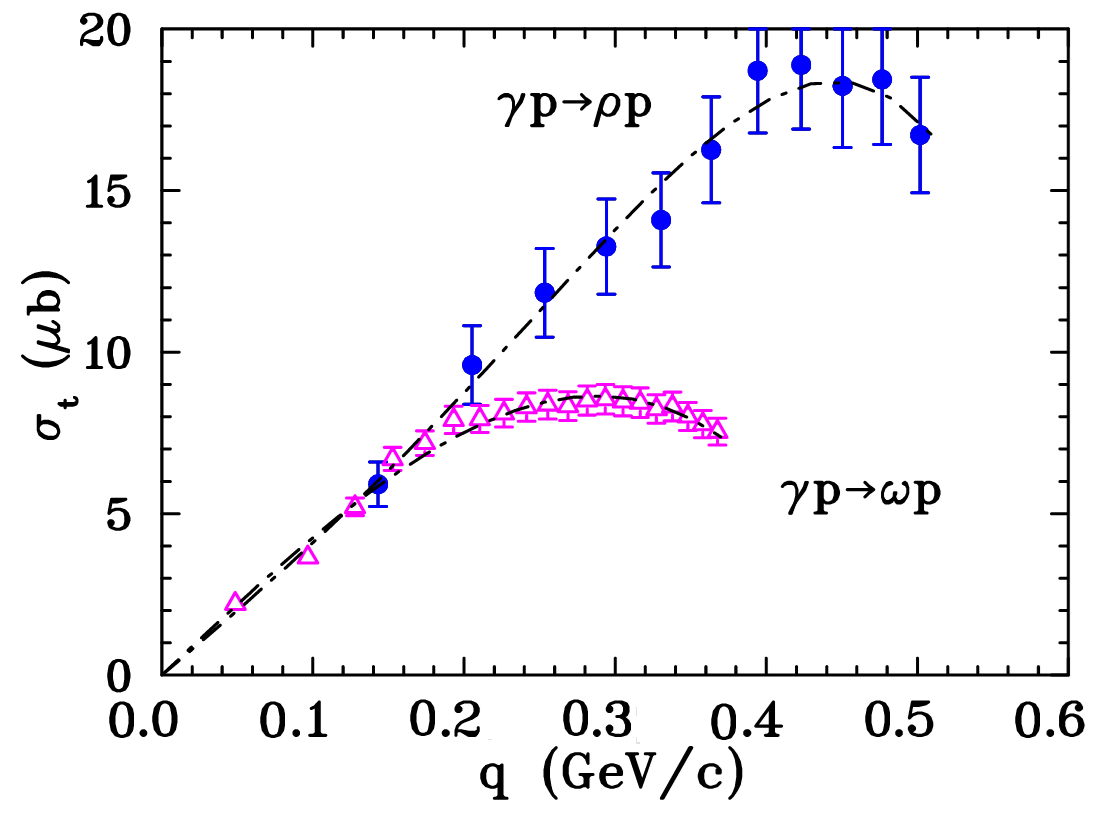}
}
\centerline{\parbox{0.4\textwidth}{
\caption[] {\protect\small
The total $\gamma p \to Vp$ cross section $\sigma_t$ derived from the A2 at MAMI (magenta open triangles)~\cite{Strakovsky:2014wja} and CLAS (blue filled circles) quasi-data~\cite{CLAS:2018drk} is shown as a function of the center of mass momentum $q$ of the final-state particles. Black dash-dotted curves are the polynomial fits (Eq.~(\ref{eq:eq1})) of the data.
} 
\label{fig:fig2} } }
\end{figure}

To avoid an effect of a broad width of the $\rho$-meson ($\Gamma_\rho = 147.4\pm 0.8~\mathrm{MeV}$~\cite{ParticleDataGroup:2024cfk}) in the treatment of the threshold data, we are working far away from the threshold (our minimal momentum of the $\rho$ is $q_{min} = 143~\mathrm{MeV/c}$ while for the $\omega$ case it was $q_{min} = 49~\mathrm{MeV/c}$). We assume that there are no $\rho N$ bound states below the experimental $q_{min}$, or that the effect is minimal and consistent with cross section uncertainties.
Branching fraction for the decays to $\rho N$ final state for the nucleon resonances located near $\rho p$ threshold, defined with the central $\rho$ mass, is smaller than 10\%~\cite{ParticleDataGroup:2024cfk}. There is no evidence for the manifestation of the resonance-like structure in $q$-evolution of $\gamma p \to \rho p$ cross sections shown in Fig.~\ref{fig:fig2}.
Finally, in the extraction of $\rho N$ SL, we treat the $\rho$-meson as a quasi-stable object similar to $\omega$. BTW, the analysis of Wang \textit{et al.}~\cite{Wang:2022zwz} did the same. 

Remarkably, that the $\sigma_t$s for both $\rho$ and $\omega$ photoproduction below $q = 0.2~\mathrm{GeV/c}$ ($W = 1.76~\mathrm{GeV}$) are equal to each other. Then $\rho$ cross section grows rapidly (Fig.~\ref{fig:fig2}).

The best fit results $a = (3.99\pm 0.52)\times 10^{-2}~\mathrm{\mu b/(MeV/c)}$ for the $\rho$ case, which one compares to the $\omega$ case  
$a = (4.42\pm 0.14)\times 10^{-2}~\mathrm{\mu 
b/(MeV/c)}$ using analysis of the A2 Collaboration at MAMI measurements~\cite{Strakovsky:2014wja}. All figures are given in Table~\ref{tbl:tab1}. The main source of SL uncertainties came from the best fit results using Eq.~(\ref{eq:eq1}) (uncertainty of $\Delta a$).

\begin{table}[htb!]
\centering \protect\caption{{
The second column shows the minimal momentum $q_{min}$ for V in photoproduction experiments on the proton target, along with the source of the data.  
The decay $\Gamma(V\to e^+e^-)$ from PDG2024~\cite{ParticleDataGroup:2024cfk} (third column). 
The fourth column showed phenomenological Vp absolute value SLs, $|\alpha_{Vp}|$, and the sources of results: for $\omega$ by A2 Collaboration at MAMI~\cite{Strakovsky:2014wja} by ELPH~\cite{Ishikawa:2019rvz} and CBELSA/TAPS~\cite{Han:2022khg, CBELSATAPS:2015wwn} Collaborations; and for $\rho$ by CLAS Collaboration and by 
SAPHIR Collaboration~\cite{Wang:2022zwz, Klein:1996}.}
}
\vspace{2mm}
{%
\begin{tabular}{|c|c|c|c|}
\hline
Meson         & q$_{min}$    & $\Gamma(V\to e^+e^-)$ & $|\alpha_{Vp}|$     \tabularnewline
              &   (MeV/c)    &  (keV)    & (fm)       
\tabularnewline
\hline
$\rho(770)$   & 143~\cite{CLAS:2018drk} & 7.04$\pm$0.06              &
0.23$\pm$0.03
\tabularnewline
              &                       &                              &
             0.24$\pm$0.02~\cite{Wang:2022zwz}
\tabularnewline
\hline
$\omega(782)$ &  49~\cite{Strakovsky:2014wja}       & 0.60$\pm$0.02  & 
0.82$\pm$0.03~\cite{Strakovsky:2014wja} 
\tabularnewline
              &                       &                              &
             0.97$\pm$0.16~\cite{Ishikawa:2019rvz} 
\tabularnewline
              &                       &                              &
             0.811$\pm$0.019~\cite{Han:2022khg}      
\tabularnewline
\hline
\end{tabular}} \label{tbl:tab1}
\end{table}

\section{Discussion and Outlook}
\label{summary}

Using tree-level meson-baryon diagrams, the JM model
allows us to extract total cross sections for the reaction $\gamma p\to \rho p$ from CLAS data for $\gamma p\to \pi^+\pi^- p$.
It was observed that $\sigma_t (\gamma p\to \rho p)$ is close to $\sigma_t(\gamma p\to \omega p)$ below $q = 0.2~\mathrm{GeV/c}$. Then, the $\rho$p photoproduction cross section is growing rapidly.

We found that the $\rho$ nucleon SL result is by a factor of 4 smaller than the size of the hadron and the phenomenological determination of the $\omega$ nucleon SL using threshold photoproduction cross sections. Our analysis was performed using CLAS data within the JM model.

Our $\rho p$ SL, $|\alpha_{\rho p}|$, using CLAS data~\cite{CLAS:2018drk} is in good agreement with the results~\cite{Han:2022khg} using an analysis of the previous SAPHIR measurements~\cite{Klein:1996}.
Several recent phenomenological results for $\omega p$ SL, $|\alpha_{\omega p}|$, by the A2 Collaboration at MAMI~\cite{Strakovsky:2014wja} by ELPH~\cite{Ishikawa:2019rvz} and CBELSA/TAPS~\cite{Han:2022khg, CBELSATAPS:2015wwn} Collaborations are in good agreement (Table~\ref{tbl:tab1}).

Meanwhile, the dynamical coupled-channel approach addresses the analysis of the $\pi^-p \to \omega n$ was performed to determine 
$|\alpha_{\omega p}|$~\cite{Wang:2022osj}. The channel space of $\pi N$, $\pi\Delta$, $\sigma N$, $\rho N$, $\eta N$, $K \Lambda$, and $K\Sigma$ is extended by adding the $\omega N$ final state. This approach is free from the VMD model contribution. The authors reported two best-fit results for $|\alpha_{\omega p}| = -0.24 + i~0.05~\mathrm{fm}$ and $-0.21 + i~0.05~\mathrm{fm}$. Let us point out that we obtained $ReSL > ImSL$, which is natural for the threshold dynamics.

\section{Acknowledgments}
We thank Misha Ryskin for valuable comments and discussions.
This work was supported in part by the U.S.~Department of Energy, Office of Science, Office of Nuclear Physics under Awards No. DE--SC0016583 and No. DE--AC05--06OR23177.

\end{document}